\documentclass[%
reprint,
superscriptaddress,
groupedaddress,
nofootinbib,
nobibnotes,
amsmath,amssymb,
aps,
prl,
]{revtex4-2}

\usepackage[dvipsnames]{xcolor}
\usepackage{graphicx}
\usepackage{dcolumn}
\usepackage{bm}
\usepackage[breaklinks=true,colorlinks,citecolor=blue,linkcolor=red,urlcolor=blue]{hyperref}
\usepackage{braket}
\usepackage[utf8]{inputenc} 
\usepackage{mathtools}
\usepackage[bottom]{footmisc}
\usepackage{amsfonts}
\usepackage{amsmath}
\usepackage{slashed}
\usepackage{amsthm}
\usepackage{mathbbol}
\usepackage{soul}
\usepackage{booktabs}
\usepackage{pifont}
\usepackage{tikz}

\newcommand{\cmark}{\scalebox{1.2}{\ding{51}}}
\newcommand{\xmark}{\scalebox{1.2}{\ding{55}}}

\begin{document}

\title{Quantum Geometry of Moiré Flat Bands: A Bipartite Lattice Paradigm Beyond Valley Physics}

\author{Xiaoting Zhou$\,^{\hyperlink{equal}{*},\hyperlink{email1}{\dagger}}$}

\author{Yi-Chun Hung$\,^{\hyperlink{equal}{*}}$}

\author{Arun Bansil$\,^{\hyperlink{email2}{\ddagger}}$}

\affiliation{Department of Physics,\;Northeastern\;University,\;Boston,\;Massachusetts\;02115,\;USA}
\affiliation{Quantum Materials and Sensing Institute,\;Northeastern University,\;Burlington,\;Massachusetts\;01803,\;USA}
\begin{abstract}
Flat bands in moiré superlattices host a rich array of correlated and topological phases, yet their understanding remains largely rooted in the valley paradigm, which requires well-defined valley band structures, fundamentally limiting its ability to capture moiré physics in the broader landscape of materials. éHere we introduce a general mechanism that goes beyond the valley paradigm for engineering moiré flat bands and their quantum geometry by exploiting the lattice-graph topology of bipartite lattices. Focusing on twisted heterobilayers of dice and honeycomb lattices, we demonstrate that sublattice-selective interlayer tunneling isolates zero-energy flat bands whose degeneracy is precisely tunable by the twist angle. Crucially, the hybridization between the dice-lattice flat bands and Dirac electrons generates a non-trivial Berry curvature and a quantum metric comparable to that of a Chern insulator. Our findings establish a rigorous lattice-graph-driven framework for designing flat-band quantum geometry, charting a route toward novel correlated states in oxide heterostructures, molecular lattices, and synthetic quantum matter.
\end{abstract}

\maketitle

\renewcommand{\thefootnote}{\fnsymbol{footnote}}
\footnotetext[1]{\hypertarget{equal}{These authors contributed equally.}}
\footnotetext[2]{\hypertarget{email1}{Contact author: \href{mailto:x.zhou@northeastern.edu}{x.zhou@northeastern.edu}}}
\footnotetext[3]{\hypertarget{email2}{Contact author: \href{mailto:ar.bansil@northeastern.edu}{ar.bansil@northeastern.edu}}}

\par The discovery of flat bands in moiré superlattices has driven a paradigm shift in condensed matter physics, where enhanced electron correlations and nontrivial quantum geometry lead to the emergence of unconventional states in materials \cite{Balents2020, PhysRevX.8.031089, Cao2018}.  To date, understanding these phases has relied almost exclusively on the valley paradigm \cite{Andrei2020, jiang20242database}, which elegantly captures the valley-contrasting Berry curvature and band topology in systems such as twisted bilayer graphene (TBG) \cite{doi:10.1073/pnas.1108174108} and transition metal dichalcogenides (TMDs) \cite{PhysRevLett.135.196402, calugaru2024MTMD, doi:10.1073/pnas.2021826118, Devakul2021}by mapping the moiré potential onto high-symmetry valleys in the monolayer Brillouin zone \cite{doi:10.1073/pnas.1108174108, PhysRevLett.135.196402}. However, this reliance on clean valley structures restricts the realization of moiré flat-band physics to a narrow subset of two-dimensional (2D) materials.

\par Beyond valley-based systems, many classes of moiré materials composed of monolayers lacking clear valley degrees of freedom \cite{PhysRevB.110.165411, An2021, pub.1143215478}, such as bipartite \cite{PhysRevLett.133.236401, PhysRevB.109.155159} and kagome lattices \cite{PhysRevB.111.075434}, offer an alternative route to flat bands with twist-angle-tunable multiplicities \cite{PhysRevLett.133.236401, PhysRevB.109.155159, PhysRevB.111.075434}. Unlike their valley-based counterparts, the isolation of these flat bands originates from the bipartite graph topology of the underlying lattice, rather than from valley projection, with multiplicities tunable via the twist angle, in stark contrast to the flat bands of valley-based systems. However, the quantum geometries of these highly degenerate zero-energy flat bands remain largely unexplored, severely limiting predictive control over the ensuing strongly correlated phases, underscoring the need for a systematic framework beyond the valley paradigm. 

\par Here we address this gap by constructing minimal tight-binding (TB) models for twisted bilayers and demonstrating that nontrivial flat-band quantum geometry emerges intrinsically from interlayer hybridization within a bipartite structure. Focusing on a twisted heterobilayer of dice lattice and graphene (tb-DG), we identify sublattice-selective tunneling pathways that preserve the bipartite structure while producing isolated zero-energy flat bands with twist-angle-dependent multiplicities. Our analysis reveals that interlayer tunneling not only isolates these flat bands but also endows them with nontrivial quantum geometry comparable to that of canonical Chern insulators, arising from hybridization between the intrinsic flat bands of the dice lattice and the dispersive Dirac states of graphene. As summarized in Table~\ref{table:01}, these results establish a systematic route to engineer quantum geometry and flat-band physics in bipartite moiré architectures beyond the valley paradigm.

\begin{table}[t]
\centering
\begin{tabular}{cccl}
\toprule
\bf{ System} & $\mathbf{N_{\text{flat}}}$ & \bf{ Valley} & \bf{ Quantum Geometry} \\
\midrule
{\footnotesize TBG}    & {\footnotesize 8 \cite{doi:10.1073/pnas.1108174108}} & {\footnotesize \cmark}  & \quad \quad \quad  \,\, {\footnotesize \cmark  \cite{PhysRevLett.124.167002, PhysRevLett.128.087002, PhysRevLett.122.106405, PhysRevB.99.155415}} \\
{\footnotesize TMD} & {\footnotesize 2 \cite{Devakul2021}} & {\footnotesize \cmark}         & \quad \quad \quad  \,\, {\footnotesize \cmark \cite{PhysRevLett.122.086402, Devakul2021, PhysRevLett.132.096602}} \\
{\footnotesize TBD}  & {\footnotesize $\gtrsim2/\theta^2$ \cite{PhysRevLett.133.236401}} & \xmark  & \quad \quad \quad \,\, \xmark \\
{\footnotesize tb-DG} & {\footnotesize $\gtrsim1/\theta^2$} & \xmark  & \quad \quad \quad  \,\, {\footnotesize \cmark}(this work) \\
\bottomrule
\end{tabular}
\caption{Number of low-energy flat bands $N_{\text{flat}}$ at small twist angles $\theta$ (including spin and valley degrees of freedom), presence or absence of valley structure in the low-energy spectrum, and whether or not the moiré flat bands exhibit nontrivial quantum geometry without breaking time-reversal symmetry. Results for TBG, K-valley TMDs, twisted-bilayer dice lattice (TBD), and twisted-bilayer dice lattice graphene (tb-DG) are compared.} 
\label{table:01}
\end{table}

\begin{figure}[t]
  \centering
  \centering
    \includegraphics[width=\linewidth]{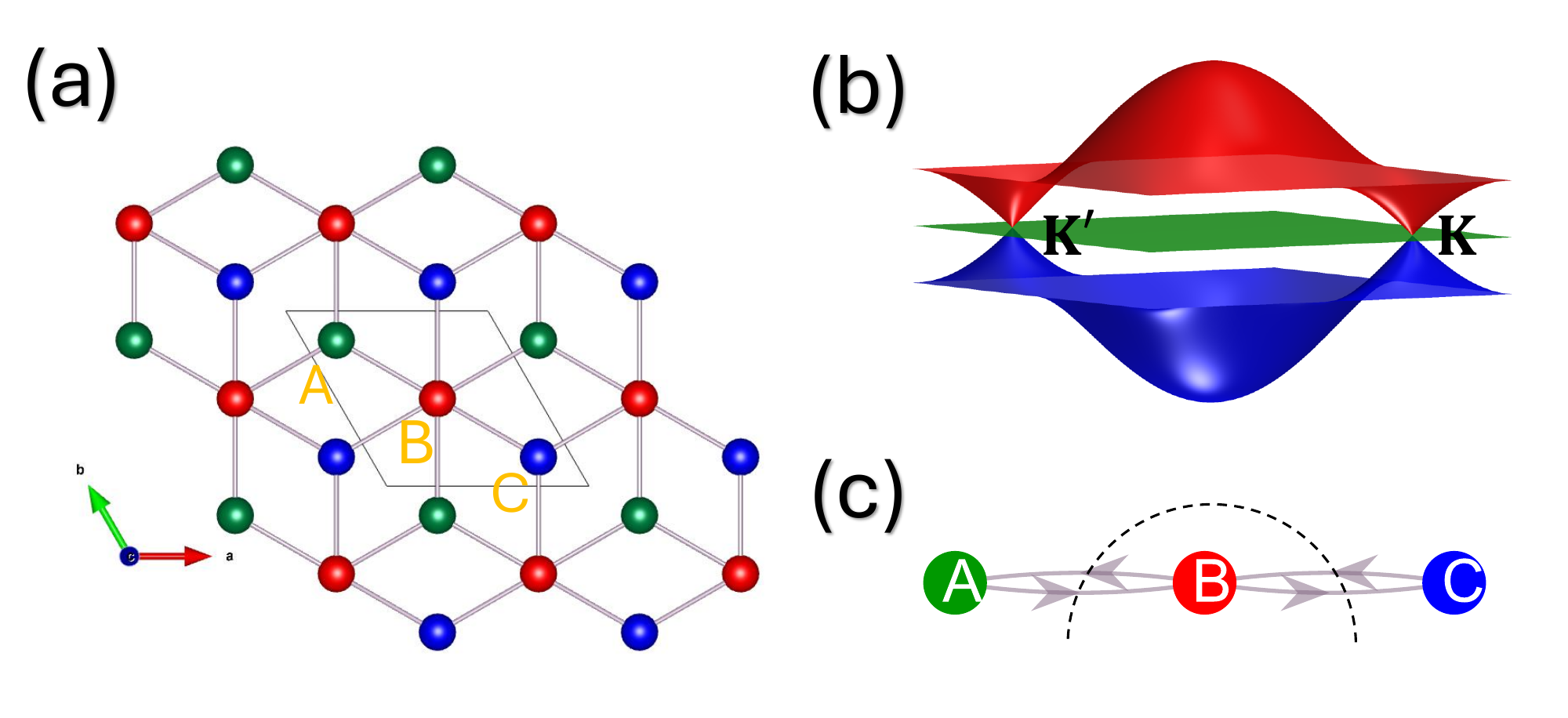}
  \caption{ (a) Dice lattice and (b) its schematic band structure. (c) The corresponding lattice graph illustrating the bipartite structure, where the partitions are separated by the black dashed line. 
  }
  \label{fig:01}
\end{figure}

\paragraph*{Dice lattice and bipartite graph topology---}The dice lattice serves as a pristine theoretical model for flat-band physics, featuring three inequivalent sublattices consisting of one sixfold coordinated hub and two threefold coordinated rim sites arranged in a geometry that extends the honeycomb lattice \cite{PhysRevB.34.5208}. The TB Hamiltonian involves hopping only between the selected nearest-neighbor (NN) sublattices (Fig.~\ref{fig:01}(a)). In the $(A,B,C)^T$ sublattice basis, the Hamiltonian is:
\begin{equation}\label{eq:dice}
    H_0^{(D)} = t\begin{pmatrix} 0 & h(\vec{k}) & 0 \\ h^*(\vec{k}) & 0 & h(\vec{k}) \\ 0 & h^*(\vec{k}) & 0 \end{pmatrix},
\end{equation}
where $h(\vec{k})=1+2e^{-i\frac{3k_y}{2}}\cos(\frac{\sqrt{3}k_x}{2})$. We set $t=-1$ in the following calculations.  A key consequence of the dice lattice is that its sublattice connectivity yields a bipartite graph topology in which the underlying lattice graph features an imbalance in the number of vertices between the two partitions (FIG.~\ref{fig:01}(c)), rendering the TB Hamiltonian noninvertible. This graph-topological characteristic guarantees the emergence of a zero-energy flat band \cite{Calugaru2022} (FIG.~\ref{fig:01}(b)), whose wave functions, arising from destructive quantum interference, reside entirely on the partition with the larger number of vertices ($A$ and $C$ sublattices) .  This follows from chiral symmetry, see Supplemental Material (SM) for details \cite{SM}\nocite{PhysRevB.74.064429, PhysRevB.75.045105, PhysRevResearch.4.043151, twist_square_2021}. The pristine flat band of the dice lattice lacks pseudospin texture and exhibits a trivial quantum geometry due to the preservation of both time-reversal and two-fold rotation symmetries.

\begin{figure}[t]
  \centering
  \centering
    \includegraphics[width=\linewidth]{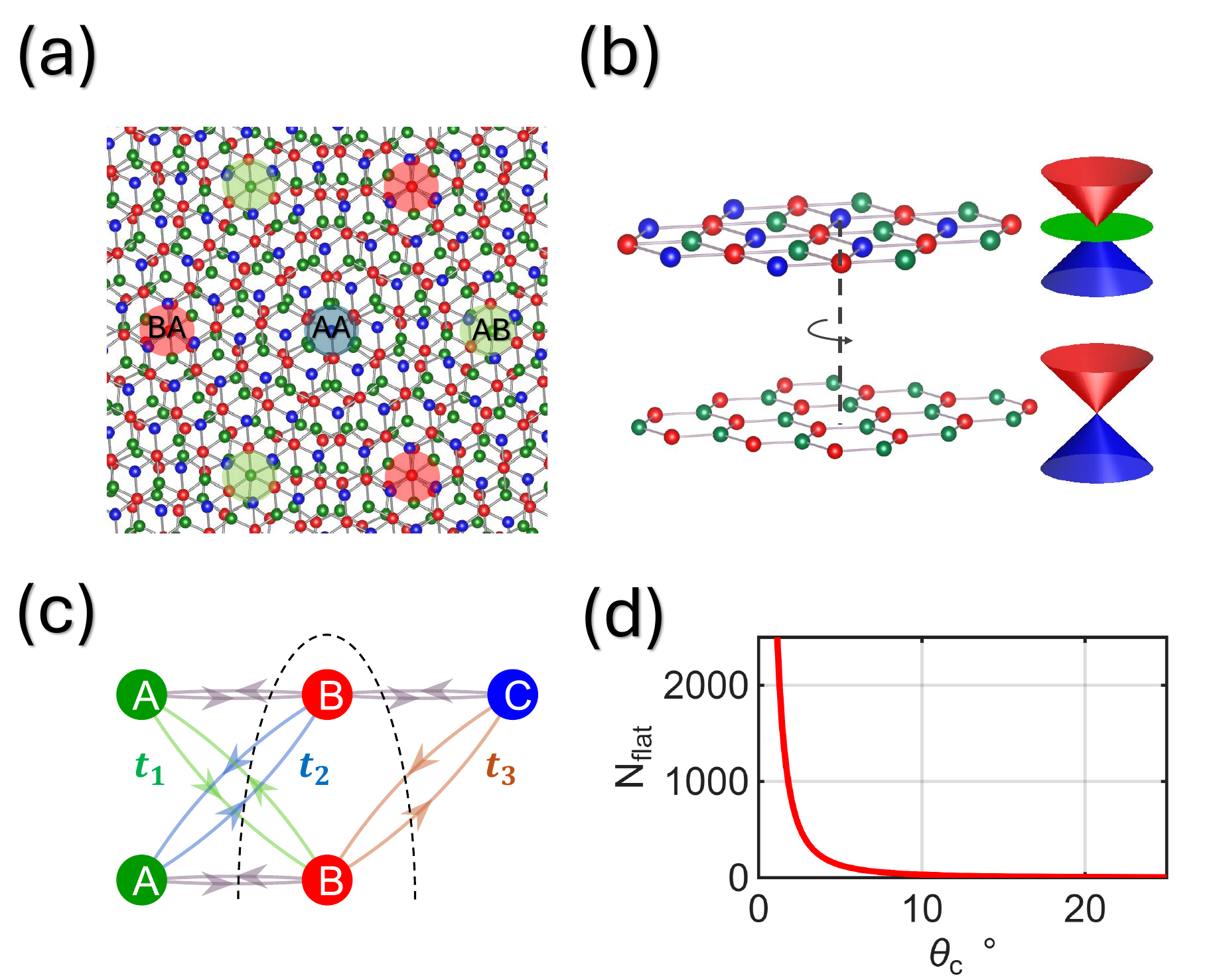}
  \caption{ (a) Top and (b) side view of the tb-DG structure at $\theta_c\approx9.43^\circ$, where the high-symmetry stacked regions are highlighted in different colors. (c) The corresponding lattice graph reveals a bipartite structure, with partitions separated by the black dashed line. Blue, green, and orange hoppings denote distinct interlayer tunnelings ($t_1$, $t_2$, $t_3$). (d) Number of zero-energy flat bands $N_{\text{flat}}$ in tb-DG lattice as a function of $\theta_c$.
  }
  \label{fig:02}
\end{figure}

\paragraph*{Sublattice-selective tunneling in tb-DG---} We model tb-DG by removing the $C$ sublattice from one layer of an AA-stacked bilayer dice lattice, aligning the honeycomb hexagon center with the dice $C$ site, and introducing a relative twist about this center. The resulting moiré pattern (Fig.~\ref{fig:02}(a)) preserves $C_{3z}$ symmetry but lacks $C_{2z}$, hence $C_{2z}T$ symmetry, which enables the emergence of a nontrivial Berry curvature.

To capture the Dirac fermion structure of graphene \cite{RevModPhys.81.109}, we retain only NN hopping on the honeycomb lattice. The global bipartite structure of the lattice graph is preserved \cite{PhysRevLett.133.236401} by introducing sublattice-selective interlayer tunneling terms (Fig.~\ref{fig:02}(c)). The resulting low-energy states are analyzed using the coupled TB Hamiltonian:

\begin{align}
    H_{(\delta_1,\delta_2,\delta_3)} = & H_{0,-\frac{\theta_c}{2}}^{(D)} + H_{0,\frac{\theta_c}{2}}^{(G)} \notag
    \\ & + t_z\sum_{\alpha,\beta}e^{-\lambda(\frac{r_{ij}}{h}-1)}\hat{C}^\dagger_{i,\alpha}\hat{C}_{j,\beta} + \text{H.c.}.
\end{align}
Here, $ H_{0,\theta_c}^{(l)}$ is the monolayer TB Hamiltonian of the $l$-th layer rotated by $\theta_c$, and superscripts $(D)$ and $(G)$ distinguish between the dice lattice and graphene layers, respectively. In this bipartite-preserving configuration, the three primary interlayer tunneling amplitudes, $t_1$, $t_2$, and $t_3$, couple the sublattice pairings $(\alpha,\beta)=(A^{(D)},B^{(G)})$, $(B^{(D)},A^{(G)})$, and $(C^{(D)},B^{(G)})$, respectively (Fig.~\ref{fig:02}(c)). We classify bipartite-preserving tunneling configurations using a triplet  $(\delta_1,\delta_2,\delta_3)$, where the binary index  $\delta_i \in \{0, 1\}$  denotes the absence or presence of the $i$-th tunneling type.
Interlayer tunneling strength is set to $t_z = 0.1|t|$, with a dimensionless decay length of $\lambda = 20$ to model the short-range nature of the van der Waals interaction. Spatial dependence is governed by $r_{ij}\equiv\|\mathbf{r}_i-\mathbf{r}_j\|$, where $\mathbf{r}_i$ denotes the position of the $i$-th sublattice and $h$ the interlayer distance. To maintain computational tractability, we retain only interlayer tunneling with values exceeding $E_{\text{cut}}^{\text{(int)}} =10^{-3}|t|$. Lowering the threshold to $E_{\text{cut}}^{\text{(int)}} = 10^{-6}|t|$ is found to yield negligible changes to the band structure. Operator $\hat{C}_{i,\alpha}$  represents the annihilation of an electron at the  $i$-th sublattice of type $\alpha$. Without loss of generality, we assume equal hopping strengths for both the dice and graphene lattices. Our analysis focuses on commensurate twist angles  $\theta_c = \theta_{c,m}$ and $60^\circ - \theta_{c,m}$, where $\theta_{c,m} = \cos^{-1} [ (3m^2 + 3m + 1/2)/(3m^2 + 3m + 1)]$ with $m \in \mathbb{N}$ \cite{PhysRevLett.99.256802}; see SM for details \cite{SM}.

\begin{figure}[t]
  \centering
  \centering
    \includegraphics[width=\linewidth]{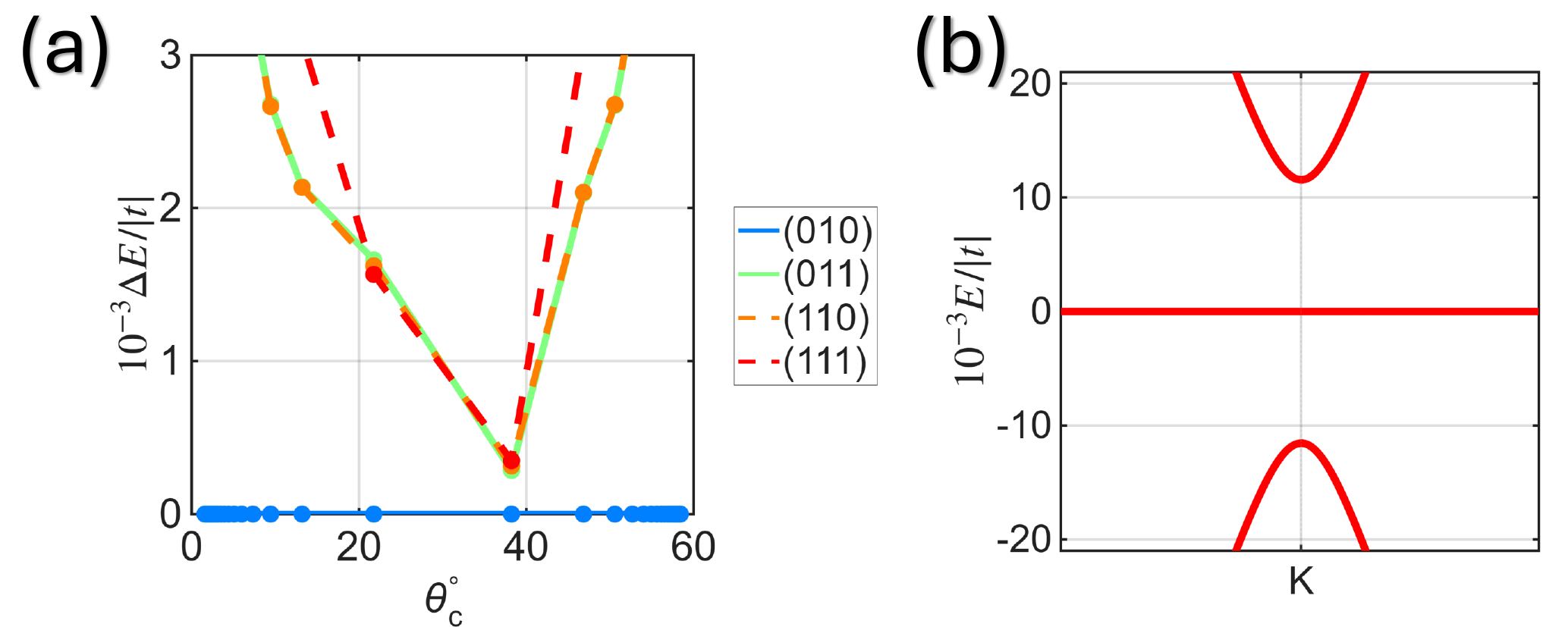}
  \caption{ (a) Band gaps between the flat bands and higher-energy bands as functions of $\theta_c$ for selected types of itnerlayer tunneling, ranging from $\theta_c\approx58.39^\circ$ to $\theta_c\approx1.61^\circ$. (b) Band structure of tb-DG at a representative twist angle of $\theta_c\approx3.15^\circ$ near the $K$ points with (111)-type interlayer tunneling.}
  \label{fig:BS}
\end{figure}

\paragraph*{Electronic structure and high degeneracy---}Under bipartite-preserving configurations, the tb-DG exhibits flat bands rigorously pinned at zero energy (FIG.~\ref{fig:BS}(b)), comprising exactly one-fifth of the total number of bands (FIG.~\ref{fig:02}(d)). This arises from the same bipartite-graph-theoretic mechanism that protects the monolayer dice flat band\cite{Calugaru2022, SM}. 

Crucially, only the interlayer tunneling configurations (110), (011), and (111) isolate the zero-energy flat bands from dispersive higher-energy states across all twist angles (FIG.~\ref{fig:BS}(a)). This can be understood from a schematic tb-DG Hamiltonian written in the basis $(A^{(D)},B^{(D)},C^{(D)},A^{(G)},B^{(G)})$:
\begin{equation}
    H_s = \begin{pmatrix} 0 & -h(\vec{k}) & 0 & 0 & t_1 
         \\ -h^*(\vec{k}) & 0 & -h(\vec{k}) & t_2 & 0 
         \\  0 & -h^*(\vec{k}) & 0 & 0 & t_3
         \\ 0 & t_2 & 0 & 0 & -h(\vec{k})
         \\ t_1 & 0 & t_3 & -h^*(\vec{k}) & 0
         \end{pmatrix}.
\end{equation}
Our focus is on achieving band separation at $K$ and $K'$ points,  where the dice-lattice monolayer exhibits degeneracies between the zero-energy flat band and the dispersive bands.  The number of flat bands as a function of the twist angle can be estimated by the condition of having commensurate twist angles, yielding  $N_{\text{flat}}=1/[2(1-\cos(\theta_c))]$. At $\vec{k}=K, K'$ (where $h(\vec{k})=0$), only the specific configurations of the interlayer tunneling, namely (110)-, (011)-, and (111)-types, preserve the generic rank of the system's Hamiltonian, $\operatorname{rank}(H_s)$. This constraint prevents the emergence of additional zero modes and ensures strict energetic isolation between the flat bands and the higher-energy dispersive bands.  This reveals a fundamental conceptual shift: zero-energy flat bands rooted in the bipartite graph topology lie beyond the conventional valley-scattering framework commonly employed in moiré physics. Consequently, even valley-dominated materials, such as graphene, require a broader theoretical paradigm when coupled to such bipartite structures. 

\begin{figure}[t]
  \centering
  \centering
    \includegraphics[width=\linewidth]{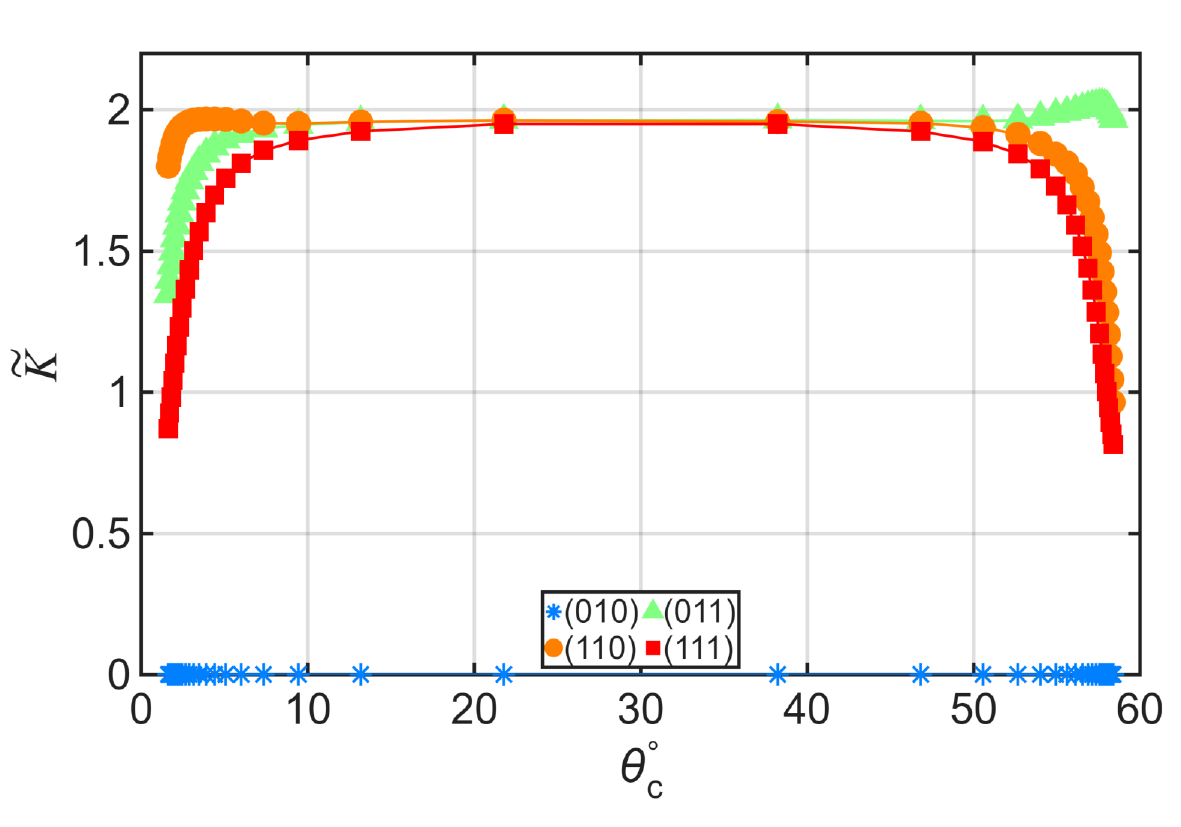}
  \caption{ The lower bound of quantum weight $\tilde{K}$ of tb-DG with various $\phi_H$ values for selected types of interlayer tunneling terms as function of $\theta_c$, ranging from $\theta_c\approx58.39^\circ$ to $\theta_c\approx1.61^\circ$.
  }
  \label{fig:Q_weight}
\end{figure}
\begin{figure}[t]
  \centering
  \centering
    \includegraphics[width=\linewidth]{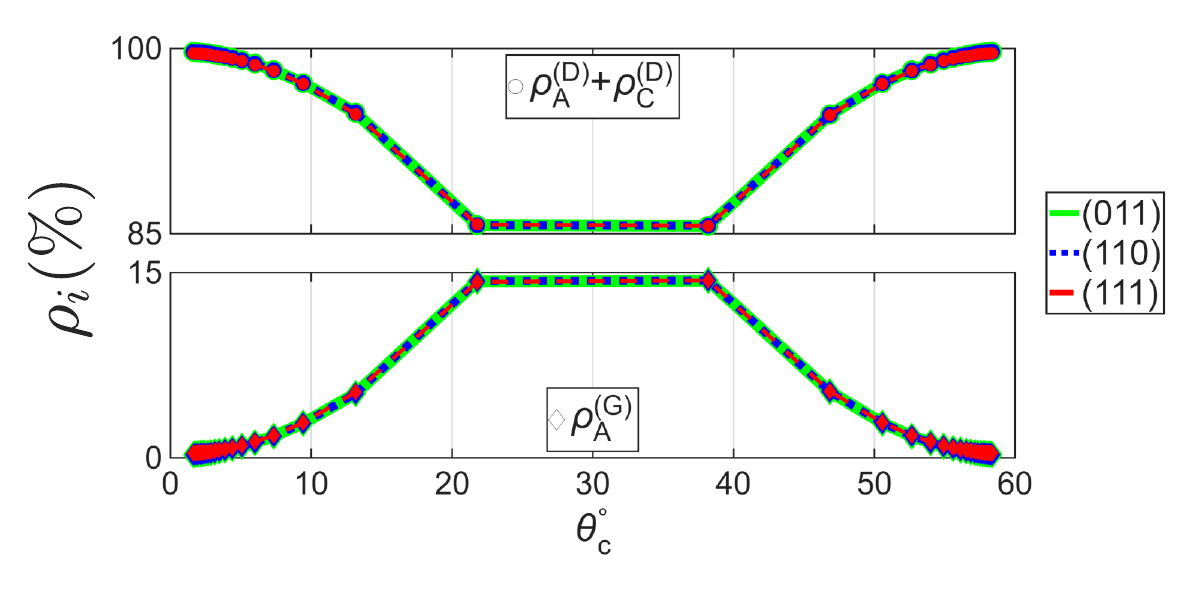}
  \caption{ Wave function compositions of isolated flat bands at the $K$ points in tb-DG with (011)-, (110)-, and (111)-types of interlayer tunneling terms, as functions of $\theta_c$ from $\theta_c\approx58.39^\circ$ to $\theta_c\approx1.61^\circ$. $\rho_{i}^{(l)}$ denotes the partial charge on sublattice $i$ of layer $l$ ($l = D, G$ for dice lattice and graphene, respectively).
  }
  \label{fig:K_comp}
\end{figure}

\begin{figure}[h]
  \centering
  \centering
    \includegraphics[width=\linewidth]{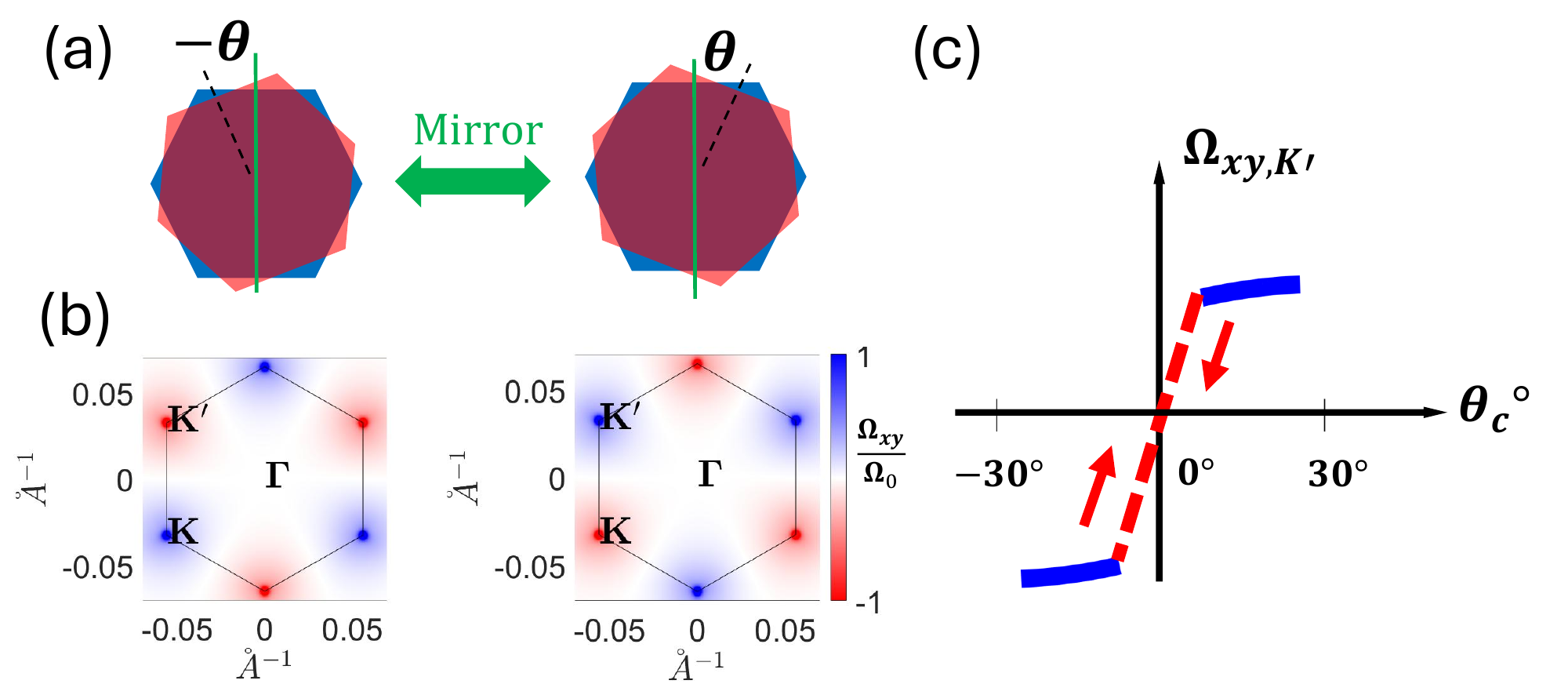}
  \caption{ (a) Twisted bilayers at $\theta_c$ and $-\theta_c$ are related by a mirror reflection. (b) Berry curvature distribution of the isolated flat bands scaled by $\Omega_0=10^{4}$ in tb-DG with (1,1,1)-type interlayer tunneling at $\theta_c\approx\pm2.13^\circ$. (c) The Berry curvature at the $K$ points $\Omega_{xy,K}$ changes sign upon mirror reflecting tb-DG with $\theta_c$ to $-\theta_c$. As $\theta_c$ tunes to $0^\circ$, $\Omega_{xy,K'}$ vanishes, leading to the decreased $\tilde{K}$ shown in FIG.~\ref{fig:Q_weight}. Similar physics applies to the systems at $\theta_c$ near $60^\circ$.
  }
  \label{fig:K_schematic}
\end{figure}

\paragraph*{Quantum geometry and Berry curvature genesis---}  The moiré effect in tb-DG extends beyond simple band folding; it actively hybridizes the electronic structures of the two layers, generating flat bands with greatly enhanced quantum geometry. We quantify this through the lower bound of quantum weight $\tilde{K}$, which establishes a rigorous lower bound for the flat-band quantum metric: 
\begin{align}
\tilde{K} \equiv  \, 2\pi\int d[\mathbf{k}] |\Omega_{xy}(\mathbf{k})|, 
\end{align}
where $d[\mathbf{k}]=d^2\mathbf{k}/(2\pi)^2$, and $\Omega_{xy}$ is the Berry curvature. Remarkably, across nearly all interlayer tunneling regimes, the Berry curvature concentrates near the points $K$ and $K'$, with opposite signs due to time-reversal symmetry. This yields a large quantum weight of $\tilde{K}\sim\mathcal{O}(1)$ (FIG.~\ref{fig:Q_weight}), comparable to that of Chern insulators.

In contrast to the trivial Berry curvature of the pristine dice lattice, tb-DG develops nontrivial quantum geometry through hybridization between the dice-lattice flat band and graphene Dirac states. Dictated by the underlying bipartite structure, the moiré flat bands are confined to the $A^{(D)}$, $C^{(D)}$, and $A^{(G)}$ sublattices, such that the emergent quantum geometry is driven by sublattice-selective interlayer coupling. This is confirmed by analyzing the wave-function composition of the isolated flat bands with (011)-, (110)-, and (111)-types of tunneling terms at $K$ point  (FIG.~\ref{fig:K_comp}). 

To isolate this mechanism, we consider the (010)-type interlayer tunneling as a controlled theoretical limit. By restricting coupling exclusively between the  $B^{(D)}$ and $A^{(G)}$  sublattices, this configuration leaves the  $A^{(D)}$- and $C^{(D)}$ -localized flat bands fully decoupled from graphene. Consequently, interlayer hybridization is suppressed, and the flat bands retain trivial quantum geometry, with our calculations confirming $\tilde{K} = 0$ across all twist angles (FIG.~\ref{fig:Q_weight}). 
This rigorously demonstrates that targeted moiré hybridization is the fundamental engine driving this large quantum geometry.

\par As shown in FIG.~\ref{fig:Q_weight}, $\tilde{K}$ steadily decreases as the twist angle approaches $0^\circ$ and $60^\circ$, eventually rendering the flat bands geometrically trivial. This quenching of quantum geometry aligns directly with the previously noted shifts in wave function composition (FIG.~\ref{fig:K_comp}) and the delocalization of Berry curvature away from the $K$ and $K'$ points (FIG.~S2 \cite{SM}). Fundamentally, the vanishing of $\tilde{K}$ is enforced by mirror symmetries (FIG.~\ref{fig:K_schematic}). Because the moiré structures at $\pm\theta_c$ are related by a mirror reflection ($M_x$ ) which reverses the sign of $\Omega_{xy}$ \cite{SM}, the quasi-continuous evolution of the band structure at small angles dictates that $\tilde{K}$ must smoothly vanish as $\theta_c \to 0$.  We emphasize that the strict $0^\circ$ and $60^\circ$ configurations do not form moiré superlattices themselves, thus serving as asymptotic bounds for this geometric quenching rather than well-defined structural fixed points.

\paragraph*{Generalization across bipartite lattices---}The bipartite graph topology architecture developed here extends far beyond tb-DG, representing a universal framework for generating flat bands with tunable multiplicity in heterobilayers with homologous bipartite structures. We demonstrate this by considering twisted bilayers of checkerboard-Lieb (tb-CBL) lattices (FIGs.~S3 and S4 \cite{SM}), where analogous flat-band structures arise from similar sublattice imbalance mechanisms. The Lieb lattice, formed by adding edge-centered sites to a square lattice \cite{PhysRevLett.62.1201}, hosts a flat band with a triply degenerate crossing at the $M$ point. While the intrinsic next-nearest-neighbor (NNN) hopping terms in the checkerboard lattice weakly break the strict bipartite structure and introduce a finite dispersion to the otherwise pristine flat bands, the underlying graph homology sustains a robust manifold of quasi-flat bands near zero energy, accounting for a striking one-fifth of the total spectral weight (FIGs.~S5 and S6 \cite{SM}). Importantly, however, the band-touching in the checkerboard lattice resides at the $M$ point \cite{PhysRevLett.103.046811} (FIGs.~S3 and S4 \cite{SM}). Because $M$ is a time-reversal invariant momentum point, the emergent Berry curvature is strictly quenched by symmetry. 

\paragraph*{Material Realization---} Because the mechanism discussed above relies on the topology of the underlying lattice graph rather than specific chemical orbitals, it offers a highly robust pathway for experimental realization across diverse physical platforms. Beyond the pvan der Waals candidates like MXenes \cite{Gogotsi2019} and MSenes \cite{PhysRevB.111.L041404}, appropriate bipartite connectivity can be engineered in oxide superlattices (e.g., SrTiO$_3$/SrIrO$_3$/SrTiO$_3$ \cite{doi:10.7566/JPSJ.87.041006} and LaAlO$_3$/SrTiO$_3$(111) \cite{PhysRevLett.111.126804}), atomic manipulation platforms (CO on Cu(111) \cite{Tassi_2024, Gomes2012, Khajetoorians2019}), and pristine topological networks like YCl \cite{geng2025YCldice}. Furthermore, highly tunable metal- and covalent-organic frameworks \cite{doi:10.1021/acs.accounts.0c00652, D1MH00935D, Jiang2021_ACS, D3SC06367D} offer bespoke lattice design. The efficacy of our graph-theoretic approach encompasses synthetic metamaterials for analog realizations in photonic \cite{Lieb_photonic}, acoustic \cite{10.1063/5.0040804}, and magnonic platforms \cite{Centala2023} as well as cold-atom arrays \cite{PhysRevB.73.144511, PhysRevA.80.063603}.

\paragraph*{Discussion---}By directly anchoring moiré flat-band physics to the lattice-graph topology of bipartite lattices, we establish a universal framework for engineering enhanced quantum geometry, independent of intrinsic monolayer valley structures. This architectural blueprint extends far beyond the putative van der Waals materials and suggests new pathways for material realization in oxide superlattices, macroscopic molecular arrays like CO on Cu(111), and highly tunable metal-organic frameworks \cite{doi:10.7566/JPSJ.87.041006, Tassi_2024, doi:10.1021/acs.accounts.0c00652}, among other possibilities. 

\par Our study reveals fundamental differences between coupling spin-$1/2$ Dirac cones and coupling spin-$1/2$ with spin-$1$ systems. In conventional moiré platforms like twisted bilayer graphene, low-energy physics is elegantly captured by the hybridization of Dirac quasiparticles \cite{doi:10.1073/pnas.1108174108}. However, introduction of a spin-$1$ Dirac cone breaks down this continuum-model paradigm. The emergent flat bands and their robust quantum geometry do not arise from perturbative quasiparticle interference but are instead rigorously enforced by the underlying lattice-graph topology. Destructive interference here is not a low-energy approximation but an exact topological consequence of the lattice graph, establishing a fundamentally new regime in which quantum geometry is dictated by graph-theoretic connectivity rather than by the hybridization of itinerant Dirac fermions.

\par Beyond the strict bipartite limit, our framework provides a highly controllable foundation for topological valley responses. Incorporating NNN hopping in the dice lattice renders the flat bands quadratically dispersive near the $K$ and $K'$ points of BZ \cite{PhysRevLett.133.236401}. Under a uniform external field and in the absence of large-momentum scattering, a well-defined valley degree of freedom emerges, supporting long-lived wave packets localized at $K$ and $K'$ \cite{Chang_2008, RevModPhys.82.1959, PhysRevB.72.085110}. Coupling the sharply concentrated valley Berry curvature to symmetry-breaking perturbations therefore unlocks a robust mechanism for the valley Hall effect \cite{Schaibley2016, Ren_2016}. By transforming lattice-graph topology into a design principle, our study establishes a versatile route to engineering quantum functionalities across a broad class of moiré materials.

\paragraph*{Acknowledgement---} The work was supported by the National Science Foundation through the Expand-QISE award NSF-OMA-2329067 and benefited from the resources of Northeastern University’s Advanced Scientific Computation Center, the Discovery Cluster, the Massachusetts Technology Collaborative award MTC-22032, and the Quantum Materials and Sensing Institute (QMSI).

\bibliography{apssamp}

\setcounter{equation}{0}
\setcounter{figure}{0}
\setcounter{table}{0}

\renewcommand{\theequation}{S\arabic{equation}}
\renewcommand{\thefigure}{S\arabic{figure}}
\renewcommand{\thetable}{S\arabic{table}}
\renewcommand{\bibnumfmt}[1]{[S#1]}
\renewcommand{\citenumfont}[1]{S#1}
\newcommand{\bk}{\boldsymbol\kappa}

\newcommand{\beginsupplement}{%
  \setcounter{equation}{0}
  \renewcommand{\theequation}{S\arabic{equation}}%
  \setcounter{table}{0}
  \renewcommand{\thetable}{S\arabic{table}}%
  \setcounter{figure}{0}
  \renewcommand{\thefigure}{S\arabic{figure}}%
  \setcounter{section}{0}
  \renewcommand{\thesection}{S\Roman{section}}%
  \setcounter{subsection}{0}
  \renewcommand{\thesubsection}{S\Roman{section}.\Alph{subsection}}%
}

\clearpage
\pagebreak
\begin{widetext}
\begin{center}
\textbf{\large Supplemental Material: Quantum Geometry of Moiré Flat Bands: A Bipartite Lattice Paradigm Beyond Valley Physics}
\end{center}
\tableofcontents

\section{S1. Chiral symmetry and zero-energy flat band in special bipartite lattices}
\par This section discusses how chiral symmetries together with the bipartite lattice structures in dice lattice and tb-DG guarantee robust zero-energy flat bands. For a bipartite system with a chiral symmetry $\Gamma$, the Hamiltonian can be expressed with only off-diagonal terms:
\begin{equation}\label{eq:chiral}
    H = \begin{pmatrix} \mathbb{0}_{N\times N} & D \\ D^\dagger & \mathbb{0}_{M\times M} \end{pmatrix},
\end{equation}
where $\text{dim}(D)=N\times M$ represents hopping between the sublattices in different partitions $p_N$ and $p_M$. In the lattice graph description, $N$ and $M$ denote the number of sublattices in $p_N$ and $p_M$. The corresponding chiral symmetry is $\Gamma=\text{diag}(\mathbb{1}_{N\times N}, -\mathbb{1}_{M\times M})$. According to the results of Ref.~\cite{Calugaru2022}, $|N-M|$ number of zero-energy flat bands emerge when $N\neq M$, so that is an imbalance between the number of sublattices in two partitions. Ratio of the number of flat bands to the total number of bands is thus $(N-M)/(N+M)$. In addition, equation~\eqref{eq:chiral} indicates that the wave functions of such zero-energy flat bands must localize on sublattices in the partition with a larger number of sublattices \cite{Calugaru2022}. The dice lattice and tb-DG thus separate $A,C$ and $B$ sublattices into different partitions. Given equal sublattice counts of $N$ in the dice lattice and tb-DG, we obtain $2N-N$ and $3N-2N$ flat bands, resulting in fractions of $(2N-N)/(2N+N)=1/3$ and $(3N+2N)/(3N+2N)=1/5$ zero-energy flat bands in these two cases, respectively. Therefore, the zero-energy flat-band wave functions in the dice lattice reside only on the $A$ and $C$ sublattices, while in tb-DG, they reside only on the $A$ and $C$ sublattices of the dice layer and the $A$ sublattice of the graphene layer, independent of the interlayer tunneling type.

\section{S2. Commensurate twist angles in twisted bilayer dice and honeycomb lattices}
\par Here we discuss the selection of commensurate twist angles in twisted bilayer dice and honeycomb lattices (t-D/h). Due to the underlying dice lattice, only the $C_{3z}$ symmetry is preserved, constraining the twist angles to $\theta_c \in (0^\circ, 120^\circ)$. This range can be narrowed by noting that the structure at $\theta_c$ maps to that at $- \theta_c$ (or, equivalently, $120^\circ - \theta_c$) by a mirror reflection (e.g., $M_x$ or $M_y$), see FIG.~\ref{fig:theta}(a). Moreover, due to the $C_{6z}$ symmetry of graphene, $60^\circ \pm \delta\theta_c$ map to each other, where $\delta\theta_c$ describes the commensurate twist angle around $60^\circ$. As a result, it is sufficient to study commensurate twist angles within the range $\theta_c\in(0^\circ, 60^\circ)$. A schematic of t-D/h at $\theta_c$ near $60^\circ$ is shown in FIG.~\ref{fig:theta}(b); results for $\theta_c$ values lying outside this range can be obtained by applying a mirror operation. Without loss of generality, we do so by choosing $\theta_c = \theta_{c,m}$ for $\theta_c<30^\circ$ and $\theta_c = 60^\circ - \theta_{c,m}$ for $\theta_c>30^\circ$, where $\theta_{c,m} = \cos^{-1}[(3m^2 + 3m + 1/2)/(3m^2 + 3m + 1)]$ with $m \in \mathbb{N}$ lying in $(0^\circ, 30^\circ)$. 

\begin{figure}[t]
  \centering
  \centering
    \includegraphics[width=\linewidth]{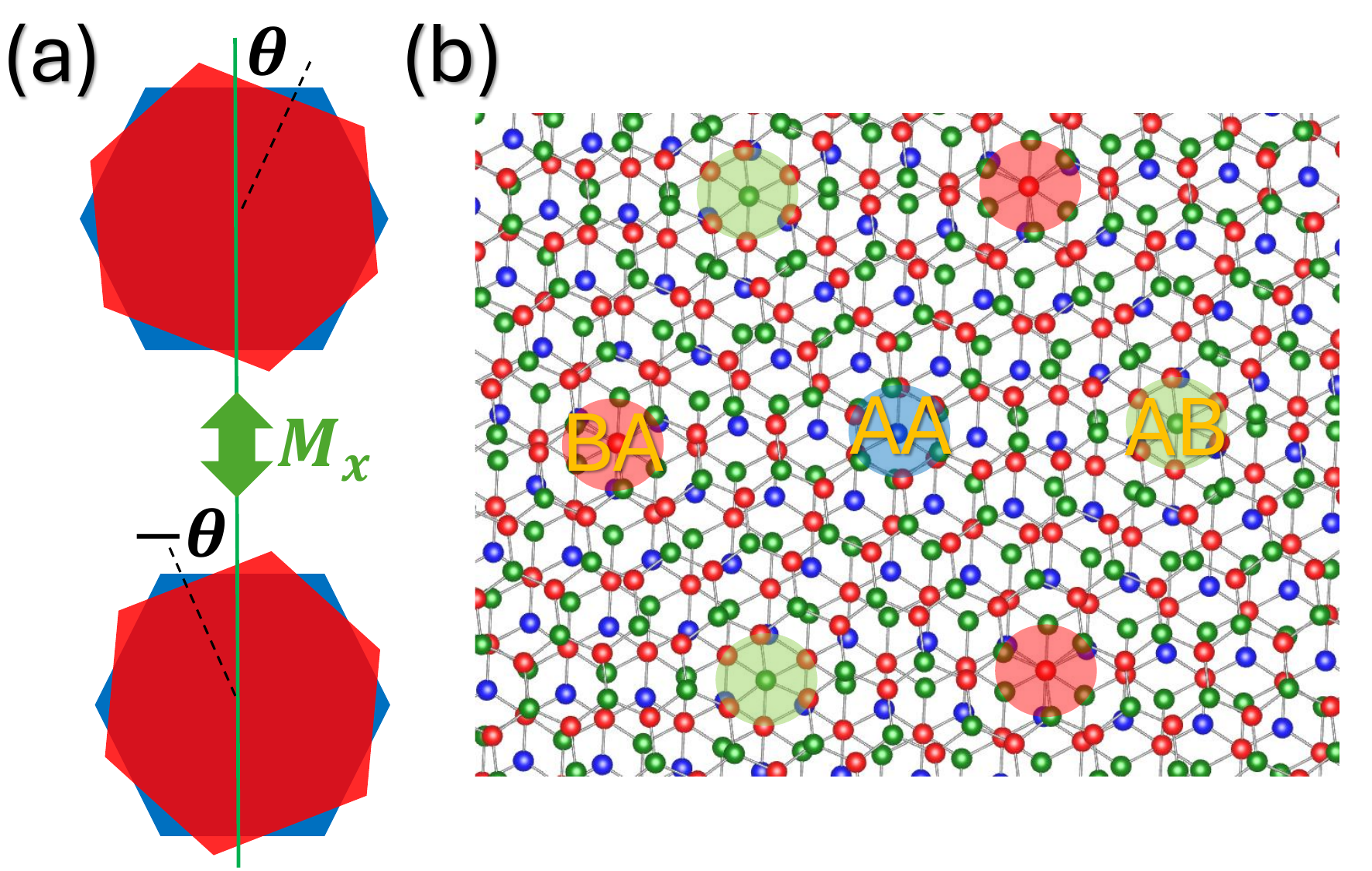}
  \caption{ (a) Twisted bilayers at $\theta_c$ and $-\theta_c$ are shown related by a mirror reflection. (b) Lattice structure of tb-DG at $\theta_c\approx69.43^\circ$, where the high-symmetry stacked regions are highlighted in different colors.
  }
  \label{fig:theta}
\end{figure}

\section{S3. Detailed electronic structure of the twisted bilayer as a function of twist angle}
\par Here we present the detailed electronic structure of tb-DG at various commensurate twist angles, including the Berry curvature distribution of the isolated flat bands at selected twist angles (FIG.~\ref{fig:curvature}). As noted in the main text, the Berry curvature of the flat bands becomes less concentrated near the $K,K'$ valleys as the twist angle decreases. Without loss of generality, we demonstrate this effect by comparing the results at $\theta_c\approx6^\circ$, $\theta_c\approx2.13^\circ$, $\theta_c\approx54^\circ$, and $\theta_c\approx57.87^\circ$ in tb-DG with $(1,1,1)$-type interlayer tunneling (FIG.~\ref{fig:curvature}). Band gaps and modified quantum weights as a function of the twist angle for various types of interlayer tunneling are shown in Figs.~\ref{fig:BS} and \ref{fig:Q_weight}, respectively.

\begin{figure}[t]
  \centering
  \centering
    \includegraphics[width=\linewidth]{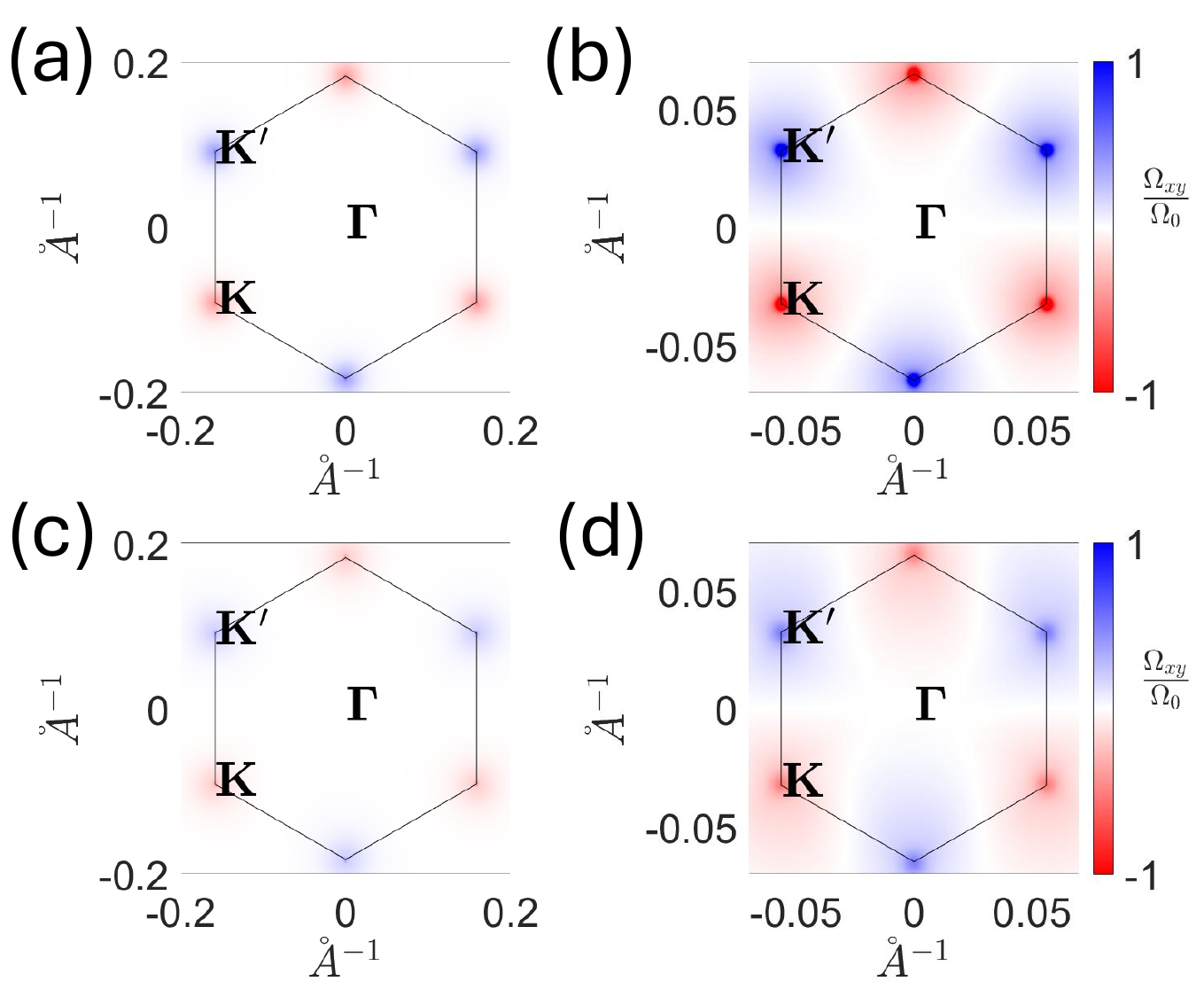}
  \caption{ Berry curvature distribution of the isolated flat bands scaled by $\Omega_0=10^{4}$ in tb-DG with (1,1,1)-type interlayer tunneling at (a) $\theta_c\approx6^\circ$, (b) $\theta_c\approx2.13^\circ$, (c) $\theta_c\approx54^\circ$, and (d) $\theta_c\approx57.87^\circ$.
  }
  \label{fig:curvature}
\end{figure}

\begin{figure}[t]
  \centering
  \centering
    \includegraphics[width=\linewidth]{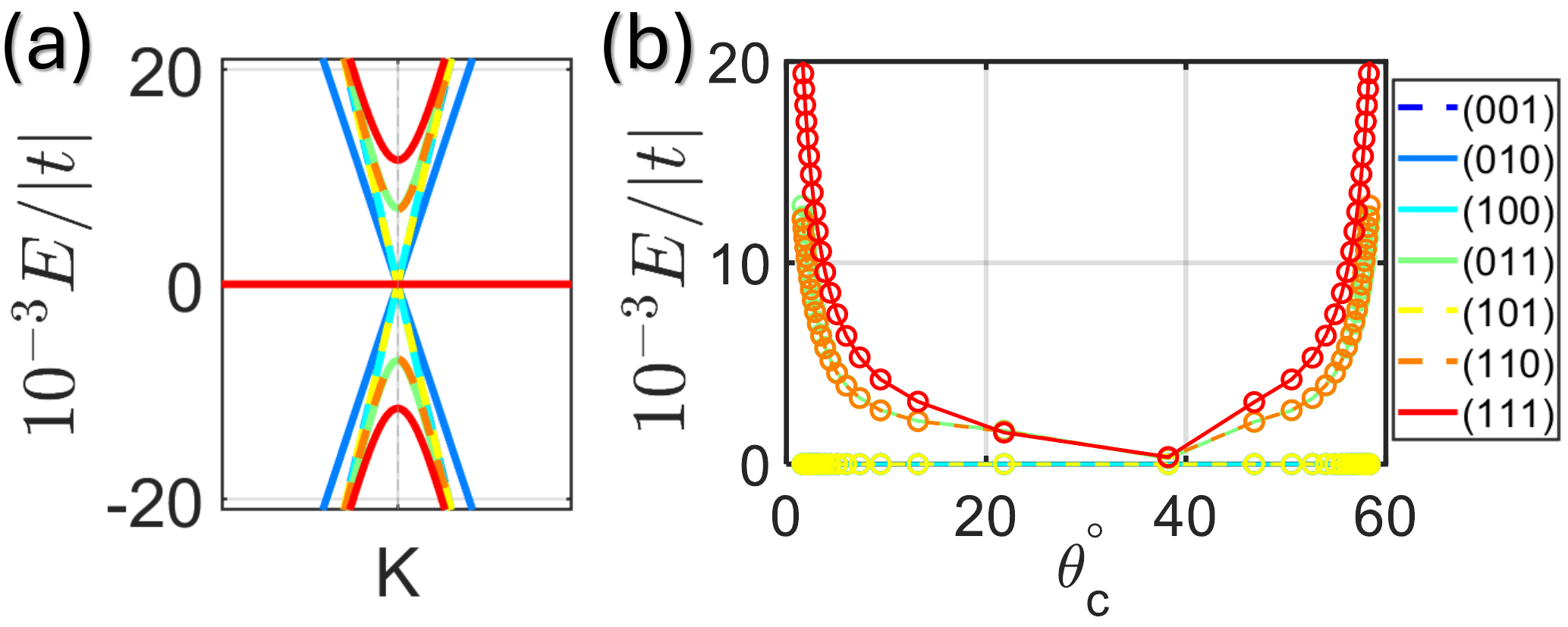}
  \caption{ (a) Band structures of tb-DG at a representative twist angle $\theta_c\approx3.15^\circ$ near the $K$ points. (b) Band gaps between the flat bands and high-energy bands as a function of $\theta_c$, ranging from $\theta_c\approx58.39^\circ$ to $\theta_c\approx1.61^\circ$.}
  \label{fig:BS}
\end{figure}

\begin{figure}[t]
  \centering
  \centering
    \includegraphics[width=\linewidth]{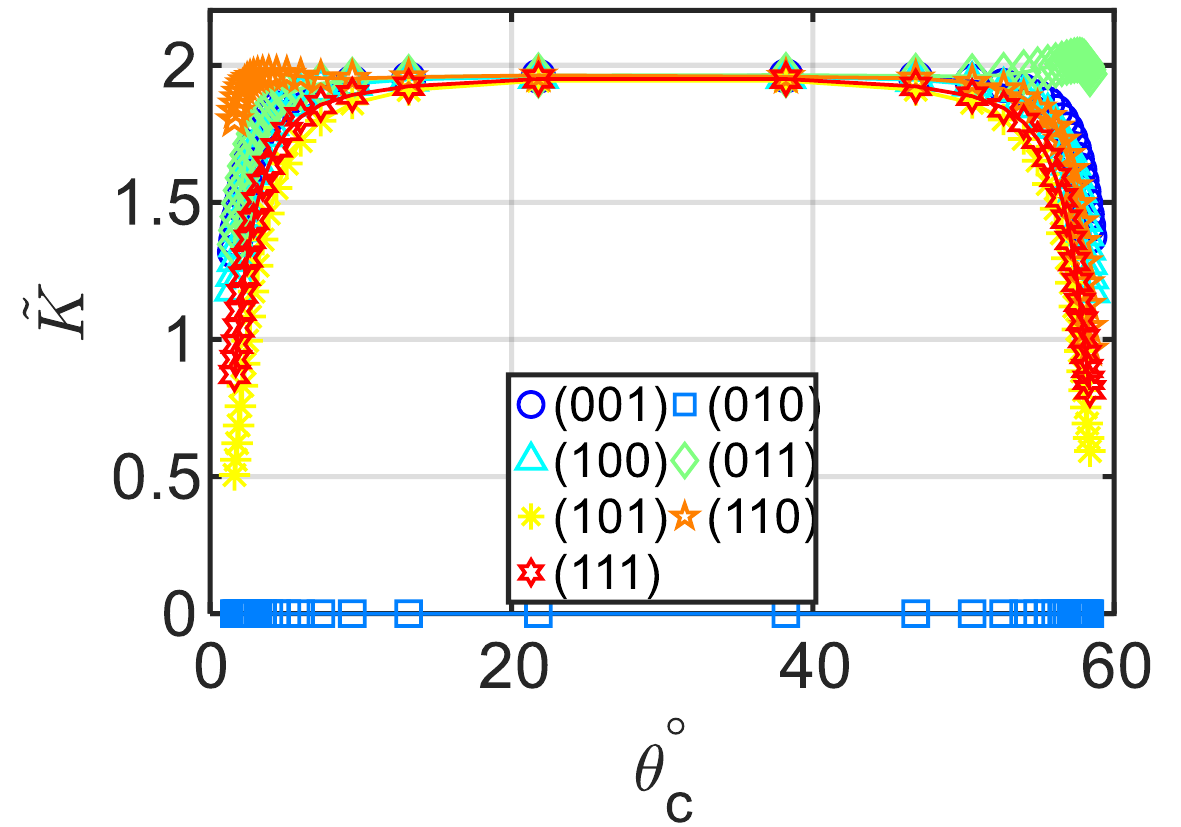}
  \caption{Modified quantum weight $\tilde{K}$ of tb-DG with various $\phi_H$ for all types of interlayer tunneling terms as a function of $\theta_c$, ranging from $\theta_c\approx58.39^\circ$ to $\theta_c\approx1.61^\circ$.
  }
  \label{fig:Q_weight}
\end{figure}

\section{S4. twisted heterobilayer of checkboard and Lieb lattices (tb-CBL)}
\subsection{S4.i Lieb lattice}
The Lieb lattice is a 2D bipartite lattice derived from the square lattice by adding one additional site at the center of each edge \cite{PhysRevLett.62.1201} (FIG.~\ref{fig:Lieb}(a)). Its TB Hamiltonian consists only of NN hopping terms, featuring a characteristic zero-energy flat band intersecting the dispersive bands at $M$ (FIG.~\ref{fig:Lieb}(b)), which, in the basis of $\big(A,B,C\big)^T$ sublattices, is:
\begin{equation}
    H_0^{(L)} = t\begin{pmatrix} 0 & 1+e^{ik_x} & 0 \\ 1+e^{-ik_x} & 0 & 1+e^{ik_y} \\ 0 & 1+e^{-ik_y} & 0 \end{pmatrix}.
\end{equation}
We chose $t=1$ in the following calculations. The Lieb lattice mirrors the graph structure of the dice lattice (FIG.~\ref{fig:Lieb}(c)), so that its zero-energy flat band has the same origin. 

\begin{figure}[h]
  \centering
  \centering
    \includegraphics[width=\linewidth]{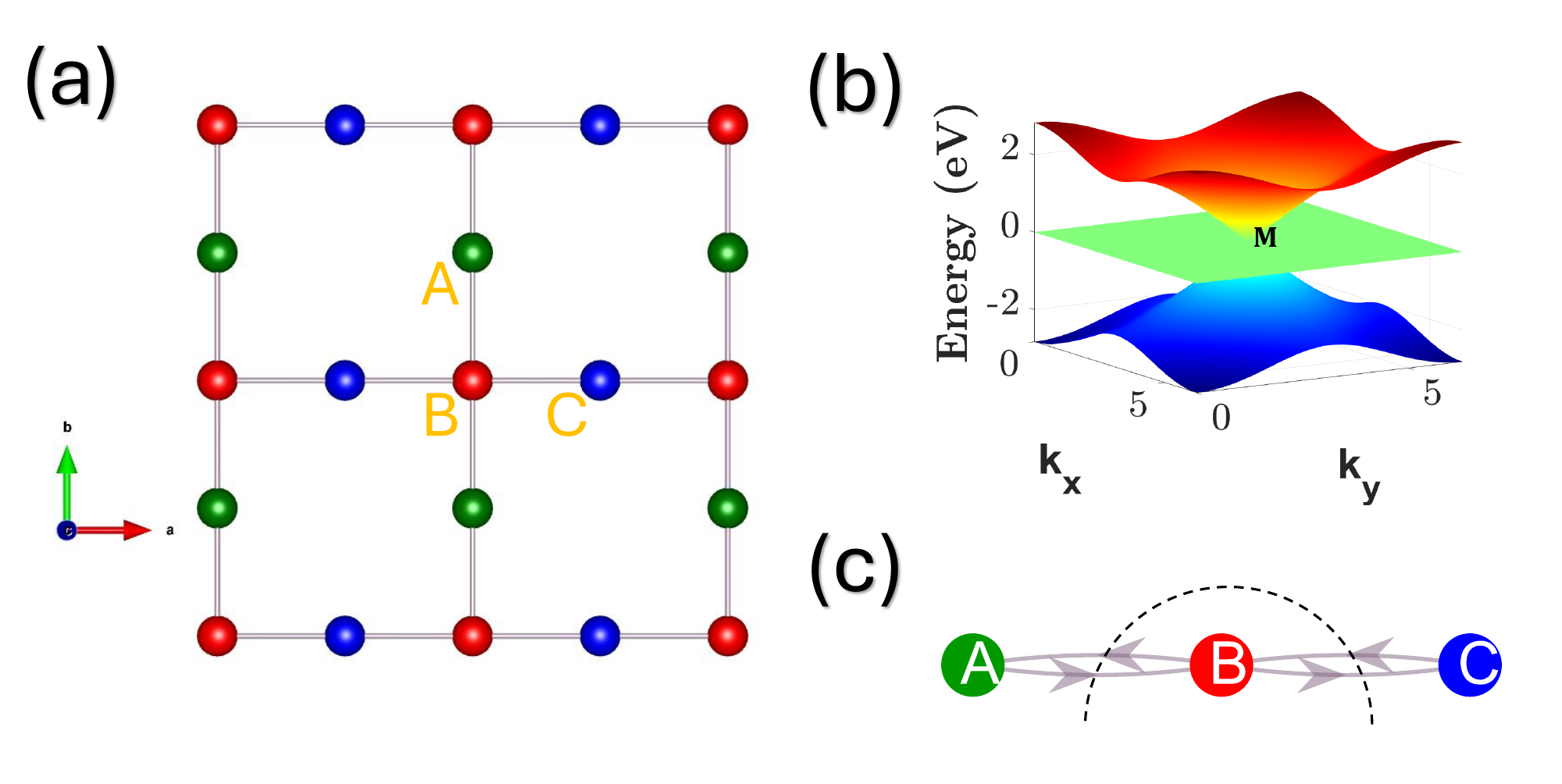}
  \caption{ (a) The Lieb lattice, and (b) its schematic band structure. (c) The lattice graph of the Lieb lattice shows a bipartite structure, where the partitions are separated by the black dashed line. 
  }
  \label{fig:Lieb}
\end{figure}
\subsection{S4.ii Checkerboard lattice (CB)}
The checkerboard lattice (CB) is a 2D square lattice with NN and NNN hopping terms on alternating plaquettes, forming a pattern reminiscent of a checkerboard (FIG.~\ref{fig:CB}(a)). While commonly studied in the context of Hubbard physics and strongly correlated phases \cite{PhysRevB.74.064429, PhysRevB.75.045105}, CB also supports non-trivial quantum geometry in its twisted bilayer. The NNN hopping in CB induces a quadratic-touching cone at the $M$ point in its monolayer (FIG.~\ref{fig:CB}(b)), supporting high Wilson loop winding number in its twisted bilayer \cite{PhysRevResearch.4.043151}. Without the NNN hopping, CB hosts a nodal band structure rather than a valley structure (FIG.~\ref{fig:CB}(c)). The TB Hamiltonian is:
\begingroup\makeatletter\def\f@size{8.5}\check@mathfonts
\def\maketag@@@#1{\hbox{\m@th\large\normalfont#1}}
\begin{equation}
    H_0^{(CB)} = t\sum_{\alpha\neq\beta\langle i,j \rangle}\hat{C}^\dagger_{\alpha,i}\hat{C}_{\beta,j} + \tilde{t}\sum_{\langle\langle i,j \rangle\rangle}\cos(2\theta_{ij})\big( \hat{C}^\dagger_{A,i}\hat{C}_{A,j} - \hat{C}^\dagger_{B,i}\hat{C}_{B,j} \big).
\end{equation}
\endgroup
Here, $\alpha,\beta\in\{A,B\}$ denote the sublattices, $i$ the lattice site, and $\theta_{ij}$ the angle between $\mathbf{r}_i-\mathbf{r}_j$ and $\hat{x}$. $\langle ... \rangle$ and $\langle\langle ... \rangle\rangle$ refer to the NN and NNN lattice sites. We choose $t=1$ and $\tilde{t}=1/2$ for the following calculations. 

\begin{figure}[t]
  \centering
  \centering
    \includegraphics[width=\linewidth]{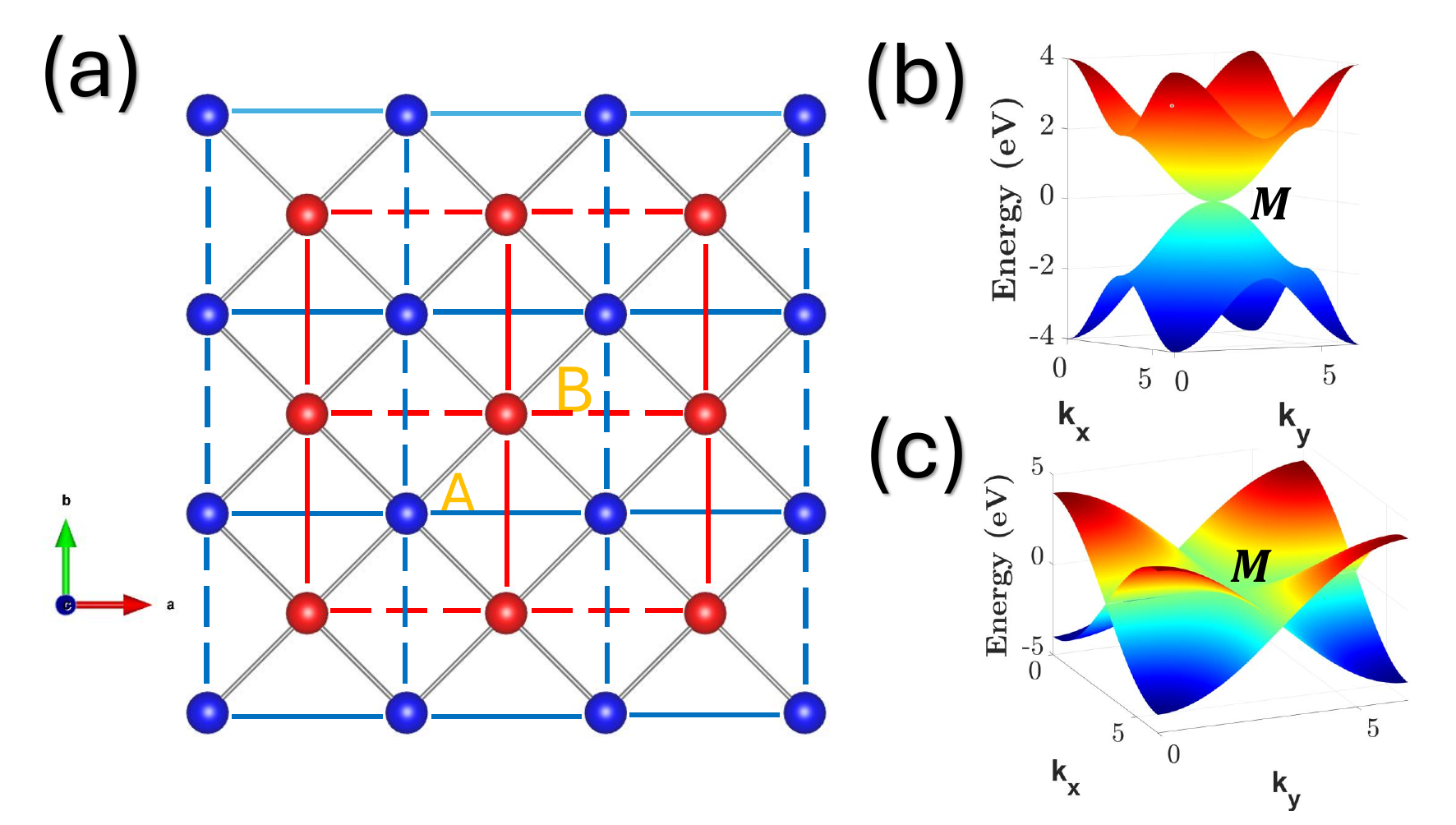}
  \caption{ (a) The heckerboard lattice (CB), where the colored solid and dashed lines mark the NNN hopping terms with positive and negative strength, respectively. (b,c) The schematic band structure of CB (b) with and (c) without the NNN hopping terms.
  }
  \label{fig:CB}
\end{figure}

\subsection{S4.iii Twisted bilayer}
\par We construct the twisted bilayer checkerboard and Lieb lattices (tb-CBL) by first aligning the $A$ sublattice of the CB layer with the $B$ sublattice of the Lieb layer, followed by a relative twist around the origin. The corresponding moiré pattern is shown in FIG.~\ref{fig:t-CB/L}(a). Without loss of generality, we apply interlayer tunneling only between the $A$ and $B$ sublattices. This yields a bipartite structure once the NNN hopping of CB is removed (FIG.~\ref{fig:t-CB/L}(b)). The corresponding full TB Hamiltonian is:
\begin{equation}
    H = H_{0,\frac{\theta_c}{2}}^{(CB)} + H_{0,-\frac{\theta_c}{2}}^{(L)} + t_z\sum_{\substack{i\in CB \\ j\in L}}e^{-\lambda(\frac{r_{ij}}{h}-1)}\hat{C}^\dagger_{i,A}\hat{C}_{j,B} + \text{H.c.}.
\end{equation}
Here, $ H_{0,\theta_c}^{(l)}$ is the monolayer Hamiltonian of the $l$th layer rotated by $\theta_c$, where $l=C,L$ denotes the checkerboard lattice at the top layer and the Lieb lattice at the bottom layer, respectively. Strength of interlayer tunneling is set to $t_z=0.1|t|$, and the dimensionless decay length is set as $\lambda=20$ to mimic the short-range nature of the van der Waals interaction. $r_{ij}\equiv\|\mathbf{r}_i-\mathbf{r}_j\|$ with $\mathbf{r}_i$ represents the position of the $j$ sublattice. $h$ is the interlayer distance. To ensure computational tractability, we retain only interlayer tunneling terms with values greater than $E_{\text{cut}}^{\text{(int)}} = 10^{-3}|t|$. $\hat{C}_{i,\alpha}$ is the annihilation operator for an electron at $i$ sublattice with type $\alpha=A,B,C$. Without loss of generality, we choose the commensurate twist angle $\theta_c$ using the relation: $\theta_c=2\tan^{-1}[1/n]$ with $n\in\mathbb{N}$ \cite{twist_square_2021}. This leads to the number of flat bands $N_{\text{flat}}=[\tan(\theta_c/2)]^{-2}+1$ (FIG.~\ref{fig:t-CB/L}(c)). The band structures of tb-CBL at $\theta_c\approx5.71^\circ$ with and without NNN hopping terms are shown in FIGs.~\ref{fig:BS_t-CBL} (a) and (b). Strong NNN hopping in the checkerboard lattice breaks the bipartite structure and results in a significant dispersion of the isolated flat bands (FIG.~\ref{fig:BS_t-CBL}(a)). Once this hopping is removed, the bipartite structure is restored in tb-CBL. Despite yielding a nodal line structure in the CB monolayer (FIG.~\ref{fig:CB}), the removal of NNN hopping still introduces isolated zero-energy flat bands with tunable number through the bipartite structure mechanism (FIG.~\ref{fig:BS_t-CBL}(b)).

\begin{figure}[h]
  \centering
  \centering
    \includegraphics[width=\linewidth]{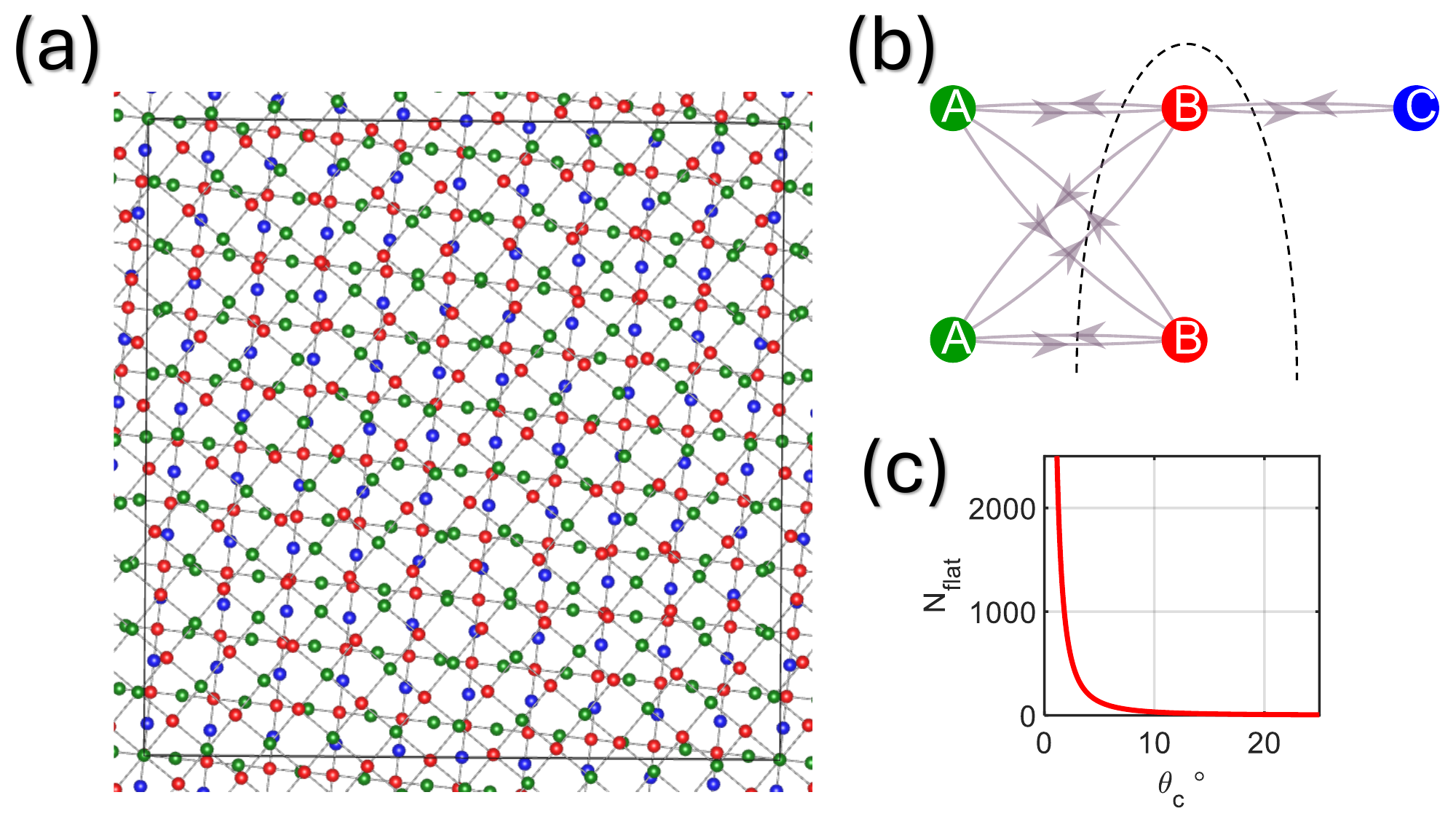}
  \caption{ (a) Lattice structure of tb-CBL at $\theta_c\approx5.71^\circ$, where the black line marks the moiré cell boundary. (b) Its lattice graph shows a bipartite structure if the NNN hopping of CB is removed, where the partitions are separated by the black dashed line. (c) Number of zero-energy flat bands $N_{\text{flat}}$ in tb-CBL as a function of $\theta_c$.
  }
  \label{fig:t-CB/L}
\end{figure}

\begin{figure}[h]
  \centering
  \centering
    \includegraphics[width=\linewidth]{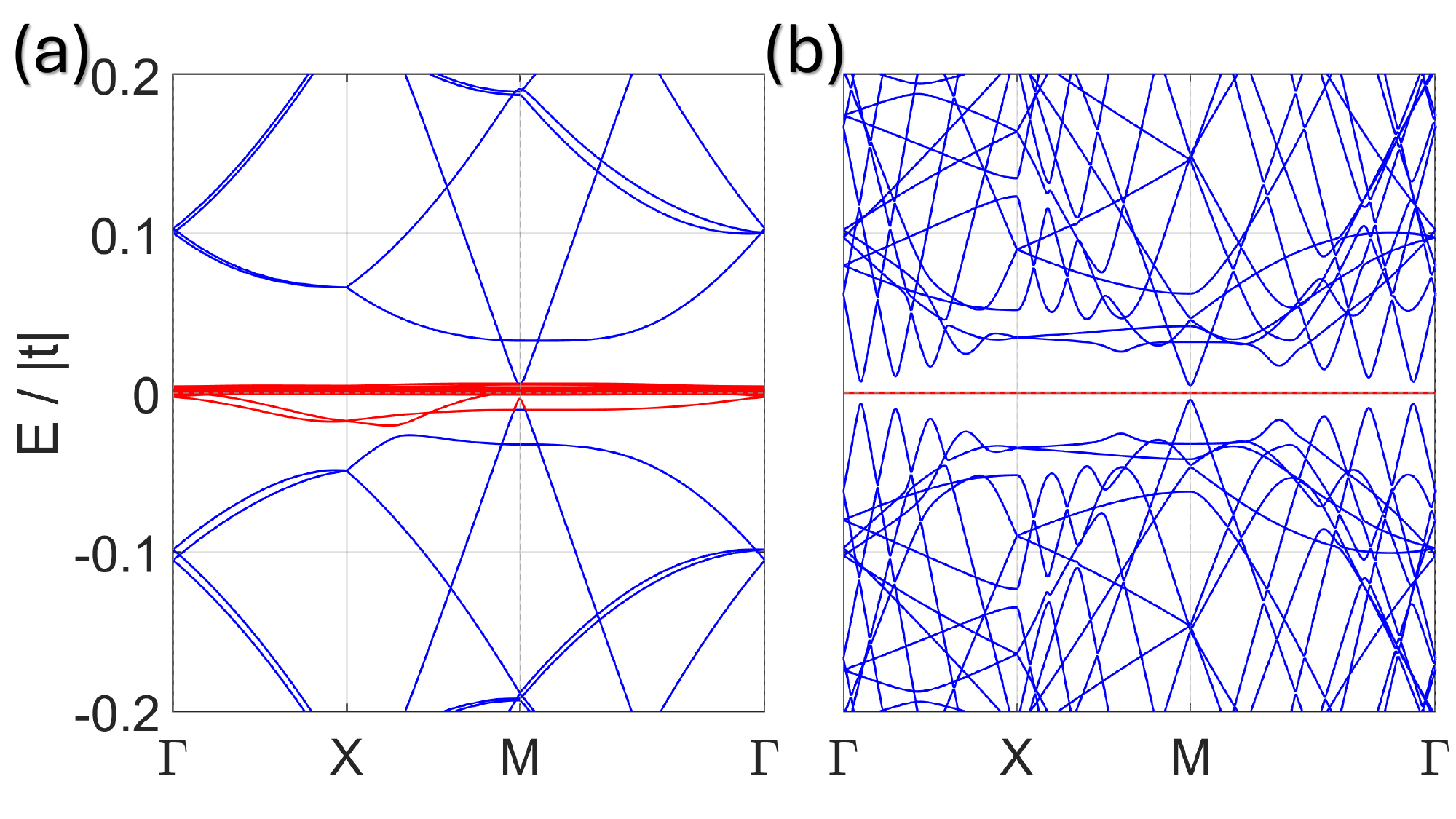}
  \caption{ Band structure of tb-CBL at $\theta_c\approx5.71^\circ$ with (a) and without (b) NNN hopping terms. The flat bands which originate here from the bipartite lattice structure are marked with red lines.
  }
  \label{fig:BS_t-CBL}
\end{figure}

\end{widetext}

\end{document}